\documentclass[twocolumn,showpacs,preprintnumbers,amsmath,amssymb,prd,letterpaper,floatfix,nofootinbib,superscriptaddress]{revtex4-1}  
\usepackage{graphicx}
\usepackage{epsfig}
\usepackage{color}
\usepackage{longtable}
\usepackage{hyperref}
\usepackage{latexsym}
\usepackage{amsmath}
\usepackage{amssymb}
\usepackage{dsfont}
\usepackage{verbatim}
\usepackage{bm}

\begin{document}


\date{\today}
\title{Evidence for $X(3872)$ from $DD^*$ scattering on the lattice}

\vspace{0.2cm}

\author{Sasa Prelovsek\footnote{sasa.prelovsk@ijs.si}}
\email{sasa.prelovsek@ijs.si}
\affiliation{Department of Physics, University of Ljubljana, Jadranska 19, 1000 Ljubljana, Slovenia}
\affiliation{Jozef Stefan Institute, Jamova 39, 1000 Ljubljana, Slovenia}

\author{Luka Leskovec}
\affiliation{Jozef Stefan Institute,  Jamova 39, 1000 Ljubljana, Slovenia}

\begin{abstract}
A candidate for the charmonium(like) state $X(3872)$ is found $11\pm 7~$MeV below the $D\bar D^*$ threshold using dynamical $N_f\!=\!2$ lattice simulation  with $J^{PC}=1^{++}$ and $I=0$. This is the first lattice simulation that establishes a candidate for $X(3872)$ in addition to  the nearby scattering states $D\bar D^*$ and $J/\psi\,\omega$, which inevitably have to be  present in  dynamical QCD. We extract large and negative  $D\bar D^*$ scattering length $a_0^{DD^*}=-1.7 \pm 0.4~$fm and the effective range $r_0^{DD^*}=0.5\pm 0.1~$fm, but their reliable determination will have to wait for a  simulation on a larger volume.  In  $I=1$ channel, only the $DD^*$ and $J/\psi\,\rho$ scattering states are found and no candidate for $X(3872)$. This is in agreement  with the  interpretation that $X(3872)$ is dominantly $I=0$, while its small $I=1$ component arises solely from the isospin breaking and is therefore absent in our simulation with $m_u=m_d$.  
\end{abstract}

\pacs{11.15.Ha,12.38.Gc,14.40.Pq, 13.25.Gv} 
\keywords{charmonium, mesonic molecule, bound state, lattice QCD} 
\maketitle 


The narrow charmonium(like) state $X(3872)$ with $m_X^{exp}\!=\!  3871.68 \pm 0.17~$MeV \cite{pdg12} has been confirmed by many experiments since its discovery  \cite{PhysRevLett.91.262001}, but its quantum numbers have been unambiguously determined to be $J^{PC}=1^{++}$ only very recently \cite{Aaij:2013zoa}.  Experimentally it is found within $1~$MeV  of the $D^0\bar D^{0*}$ threshold and has an interesting feature that it decays to isospin zero $J/\psi\,\omega$ as well as isospin one  $J/\psi\, \rho$ states. 

Theoretically it has been addressed within a great number of phenomenological models (for review see \cite{Brambilla:2010cs,Voloshin:2007dx}). The $J^{PC}=1^{++}$ charmonium channel with $I=0$ has been simulated recently also in lattice QCD  using only $\bar cc$ interpolating fields, where impressive $J^{PC}$ identification was made in \cite{Liu:2012ze}, the  continuum and chiral extrapolations were considered in \cite{Bali:2011dc,Bali:2012ua,DeTar:2012xk}, while a recent review on charmonium results from lattice is given in \cite{Mohler:2012gn}. However, $X(3872)$ has never been unambiguously identified from  a lattice simulation yet. Several simulations \cite{Liu:2012ze,Mohler:2012na,Bali:2012ua,Bali:2011dc} in fact found one state near $E\simeq m^{exp}_X$, but it was impossible to unambiguously determine whether this state is $X(3872)$ or the scattering state $D\bar D^*$, which should in principle also appear as an energy level at very similar energy  in a dynamical lattice QCD. 

There was one simulation that employed $\bar cc$ as well as $D\bar D^*$  interpolating fields, but the results could not support or disfavor the existence of $X(3872)$,  since only the lowest two energy levels in $1^{++}$ channel were extracted \cite{Bali:2011rd}.  The same holds for  \cite{Chiu:2006hd}, where only one level near the $D\bar D^*$ threshold was extracted using various four-quark interpolators.
Examples of results for near-threshold bound states  in other channels are presented in 
\cite{Beane:2013br,Yamazaki:2011nd,Beane:2003da,Liu:2012zya,Sasaki:2006jn,Albaladejo:2013aka,Liu:2008rza}.  

The purpose of our present simulation is to identify the low-lying $\chi_{c1}(1P)$ and $X(3872)$ as well as all the nearby discrete scattering levels $D\bar D^*$ and $J/\psi\, \omega$ for $I=0$. The number of energy levels near $D\bar D^*$ threshold will indicate whether we observe a candidate for $X(3872)$ or not. We search for $X(3872)$ also in the $I\!=\!1$ channel. 

In lattice QCD simulations the states are identified from discrete energy-levels $E_n$ and in principle all 
physical eigenstates with the  given quantum number appear. We employ  $J^{PC}=1^{++}$,  $I=0$ or $I=1$ and  total momentum zero. So the eigenstates are also the $s$-wave scattering states $D(\mathbf{p})\bar D^*(-\mathbf{p})$ and  $J/\psi(\mathbf{ p})V(-\mathbf{ p})$ with discrete momenta $\mathbf{p}$ due to periodic boundary conditions in space, where $V\!=\!\omega$ for $I\!=\!0$ and $V\!=\!\rho$ for $I\!=\!1$. If the two mesons do not interact then  $p\!=\!p^{n.i.}\!=\!\tfrac{2\pi}{L}|\mathbf{n}|$ and  the scattering levels appear at   $E^{n.i.}=E_1(p^{n.i.})+E_2(p^{n.i.})$. In the presence of  interaction, the scattering levels $E=E_1(p)+E_2(p)$ are shifted with respect to $E^{n.i.}$ since  momentum $p$ outside the interaction region is different from $p^{n.i.}=\tfrac{2\pi}{L}|\mathbf{n}|$. This  energy shift   provides rigorous information on the $D\bar D^*$ interaction.  Bound states and  resonances lead to levels in addition to the scattering levels and our major task is to look for these additional levels. 

Our simulation is based on one ensemble of Clover-Wilson dynamical and valence $u,d$ quarks  with $m_u\!=\!m_d$ and $m^{val}\!=\!m^{dyn}$, corresponding to $m_\pi \!=\!266(4)~$MeV. The lattice spacing  is $a\!=\!0.1239(13)~$fm, the volume is $V\!=\!16^3\times 32$ and the small spatial size $L\!\simeq\! 2~$fm  is the main drawback of our simulation considering that $X(3872)$ is probably large in size. Our exploratory results might therefore have sizable finite-volume corrections. The main purpose of this paper is, however, counting the number of lattice states near $D\bar D^*$ threshold in order to establish the existence of $X(3872)$.  

The charm quarks are treated using the Fermilab method \cite{ElKhadra:1996mp}, according to which $E-
\tfrac{1}{4}(m_{0^-}\!+\!3m_{1^-})$ are compared between lattice and experiment. 
The $m_c$  is fixed by tuning the spin-averaged kinetic mass $\tfrac{1}{4}(m_{\eta_c}+3m_{J/\psi})$ to its physical value \cite{Mohler:2012na}.  We employed the same method on this ensemble and found good agreement with experiment for conventional charmonium spectrum as well as masses and widths of charmed mesons in \cite{Mohler:2012na}. The present study also needs the following masses for our ensemble: $am_D\!=\!0.9801(10)$, $am_{D^*}\!=\!1.0629(13)$, $am_{\eta_c}\!=\!1.47392(31)$, $am_{J/\psi}\!=\!1.54171(43)$ \cite{Mohler:2012na},    $am_\rho\!=\!0.5107(40)$ and  $m_\omega\!\simeq \!m_\rho$ within errors.

The energy levels $E_n$ and overlaps $Z_i^n\equiv \langle {\cal O}_i|n\rangle$ of eigenstates $n$ are extracted from the correlation matrix 
\begin{equation}
C_{ij}(t)= \langle {\cal O}_i^\dagger (t+t_{src}) |{\cal O}_j(t_{src})\rangle=\sum_{n}Z_i^nZ_j^{n*}~e^{-E_n t}
\end{equation}
which are averaged over every second $t_{src}$. 
We choose  interpolating fields ${\cal O}_i$ that couple well to $\bar cc$ as well as the scattering states to study the system with $J^{PC}=1^{++}$ (\footnote{We employ irreducible representation $T_1^{++}$ of the lattice symmetry group $O_h$, which contains  $J^{PC}=1^{++}$ and in general also $J^{PC}\geq 3^{++}$ states, but those are at least $200~$MeV above the region of interest \cite{Mohler:2012na}.}), total momentum zero and $I=0$ or $I=1$
\begin{align}
\label{O}
O_{1-8}^{\bar cc}&=\bar c \hat M_i c(0) \qquad \qquad  (\mathrm{only}\ I=0)\\
O_1^{DD^*}&= [\bar c \gamma_5 u(0)~\bar u\gamma_i c(0) - \bar c \gamma_i u(0)~\bar u\gamma_5 c(0)] + f_I  \{u\to d\} \nonumber\\
O_2^{DD^*}\!\!\!\!&=[\bar c \gamma_5 \gamma_t u(0)~\bar u\gamma_i \gamma_t c(0) - \bar c \gamma_i \gamma_t u(0)~\bar u\gamma_5 \gamma_t c(0)]\nonumber\\
&+ f_I~  \{u\to d\}  \nonumber\\
O_3^{DD^*}&=\!\!\!\!\!\!\!\!\sum_{e_k=\pm e_{x,y,z}}\!\!\!\! [\bar c \gamma_5 u(e_k)~\bar u\gamma_i c(-e_k) - \bar c \gamma_i u(e_k)~\bar u\gamma_5 c(-e_k)] \nonumber\\
&+ f_I~ \{u\to d\}\nonumber\\
O_1^{J/\psi V}&=\epsilon_{ijk} ~\bar c \gamma_j c(0)~[\bar u\gamma_k u(0)+f_I~\bar d\gamma_k d(0)] \nonumber\\
O_2^{J/\psi V}&=\epsilon_{ijk} ~\bar c \gamma_j \gamma_t c(0)~[\bar u\gamma_k \gamma_t u(0)+f_I ~\bar d\gamma_k \gamma_t d(0) ]~,\nonumber
\end{align} 
where $f_I\!\!=\!\!1$ and $V\!\!=\!\!\omega$ for $I\!\!=\!\!0$, while $f_I\!\!=\!\!-1$ and $V\!\!=\!\!\rho$ for $I\!\!=\!\!1$.  Eight ${\cal O}^{\bar cc}$ are listed in Table X of \cite{Mohler:2012na} and polarization $i\!\!=\!\!x$ is used. Momenta are projected separately for each meson current: $\bar q_1\Gamma q_2(n)\equiv  \sum_{x}e^{i2\pi nx/L}q_1(x,t)\Gamma q_2(x,t)$ All quark fields  are smeared  $q\equiv \sum_{k=1}^{N_v}v^{(k)}v^{(k)\dagger}q_{point}$ \cite{Peardon:2009gh,Mohler:2012na} with $N_v\!=\!96$ Laplacian eigenvectors for $O^{\bar cc},~{\cal O}_{1}^{DD^*},~{\cal O}_{1}^{J/\psi V}$, and $N_v\!=\!64$ for the remaining three. The energy of $J/\psi(1)V(-1)$ is expected at least $200~$MeV above the region of interest, so the corresponding interpolator is not implemented. 

We calculate all Wick contractions (Figs. 1a, 1b of Supplemental Material) to the  correlation matrix $C_{ij}(t)$ ($13\times 13$ for $I=0$ and $5\times 5$ for $I=1$) using the distillation method \cite{Peardon:2009gh}. Certain charm annihilation  contractions  are found to be very noisy like in previous simulations. Their effect on $I=0$ charmonium states  is  suppressed due to the Okubo-Zweig-Iizuka rule, it was explicitly verified to be very small in  \cite{Levkova:2010ft} and we postpone the study of their effects to a future publication. In the present paper we present results, where $C_{ij}(t)$ contains all contractions, except for those where $c$ quark does not propagate from source to sink (results are based on contractions in Fig. 1a of the Supplemental Material).

\begin{figure*}[tb]
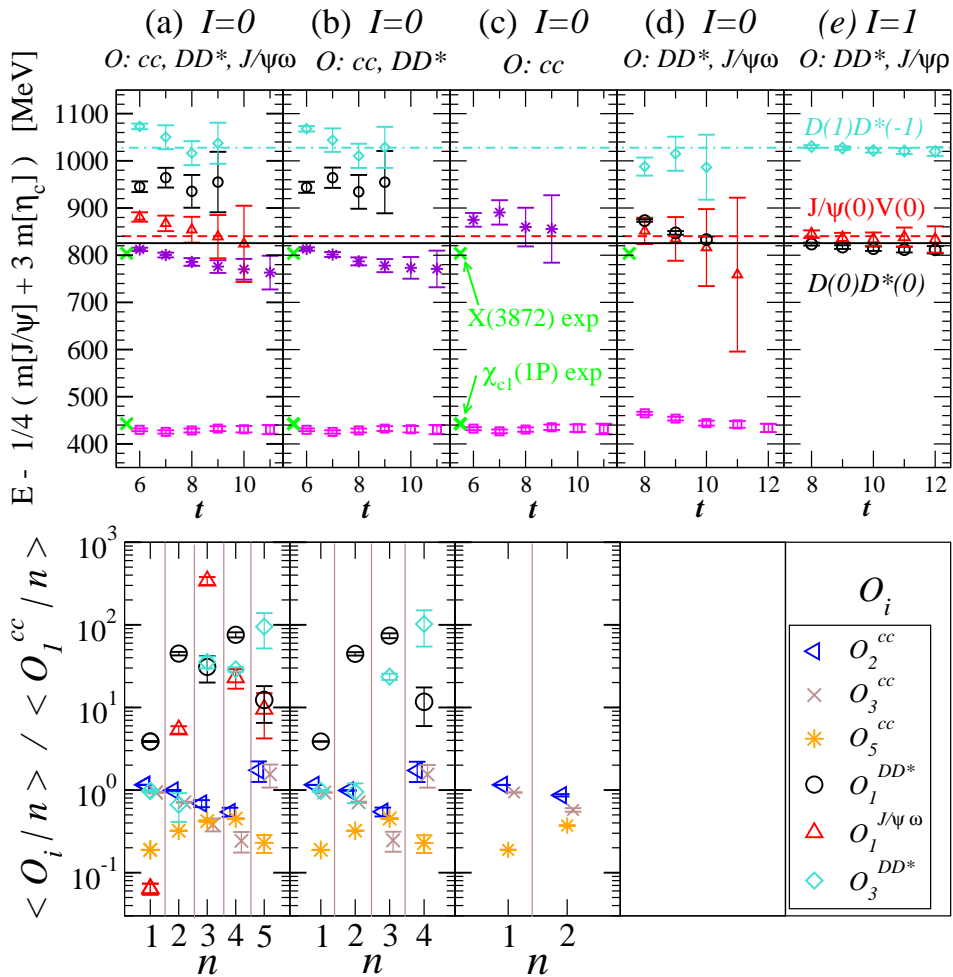

\begin{center}
\includegraphics*[width=0.7\textwidth,clip]{delMeV_T1pp.eps}
\includegraphics*[width=0.7\textwidth,clip]{rat0_T1pp.eps}
\end{center}
\caption{Upper figure: symbols represent $E_n-\tfrac{1}{4}(m_{\eta_c}+3m_{J/\psi})$ in the plateau region, where $E_{n}$ are energies of the eigenstates $n$ in the $J^{PC}=1^{++}$ channel ($n=1,2,..$ starting from the lowest state).   The choice of the interpolator basis (\ref{O})  is indicated above each plot. Dashed lines represent energies $E^{n.i.}$ of the non-interacting scattering states. Lower figure: overlaps $\langle {\cal O}_i|n\rangle$ of eigenstates $n$ (from the upper figure) with interpolators ${\cal O}_i$, all normalized to $\langle {\cal O}_1^{\bar cc}|n\rangle$.  Note that  all $\langle {\cal O}^{DD*,J/\psi \omega}_i|n\rangle/\langle {\cal O}^{\bar cc}_1|n\rangle$ depend on one (arbitrary) choice for the 
normalization of the current $\bar q_1\Gamma q_2(n)$  with a given quark smearing.  The plotted  ratios correspond to choice presented in the main text.  }\label{fig:eff}
\end{figure*}

The energies $E_n$ and overlaps $\langle O_i|n\rangle$ are extracted from the time-dependence of the correlation matrix $C_{ij}(t)$ using the generalized eigenvalue method  $C(t)u_n(t)=\lambda_n(t)C(t_0)u_n(t)$ \cite{Michael:1985ne,Blossier:2009kd}.
Results are consistent for range $2\leq t_0\leq 6$ and we present them for $t_0\!\!=\!\! 2$, when the highest level is least noisy.  The  eigenvalues $\lambda_n(t)\to e^{-E_n (t-t_0)}$ give the effective energies $E_n^{eff}(t)\equiv \log [\lambda_n(t)/\lambda_n(t+1)]\to E_n$ plotted in Fig. \ref{fig:eff}, which equal the energies $E_n$ in the plateau region.   Ratios of overlaps for state $n$ to two different interpolators \cite{Blossier:2009kd}
\begin{equation}
\frac{\langle {\cal O}_i|n\rangle}{\langle {\cal O}_j|n\rangle}=\frac{\sum_k C_{ik}(t)u_k^n(t)}{\sum_{k'} C_{jk'}(t)u_{k'}^n(t)}~
\end{equation}
evaluated at $t=8$ are also shown (they are obtained from  the full interpolator basis but  only few representative ${\cal O}_i$ are shown). 
We verify that  ratios are almost  independent of time for $6\!\leq\! t\!\leq\! 10$, indicating that our  eigenstates $n$ in Fig. \ref{fig:eff} do not change composition in time. 

The  main result of our simulation is the discrete  spectrum for $J^{PC}=1^{++}$ and $I=0,1~$ in  Fig. \ref{fig:eff}.  According to the Fermilab method for treating charm quark, we are presenting  the difference of $E$ and the spin-average $\tfrac{1}{4}(m_{\eta_c}+3m_{J/\psi})$ \cite{Mohler:2012na}, both evaluated from simulation. The horizontal lines represent energies of the non-interacting scattering states $E^{n.i.}$ on our lattice.

\vspace{0.1cm}

{\it Results for I=1:} This channel cannot contain pure $\bar cc$. The lowest three levels are $D(0)\bar D^*(0)$, $J/\psi(0)\rho(0)$ and $D(1)\bar D^*(-1)$ and we verify that their overlaps are indeed largest with ${\cal O}_{1,2}^{DD^*}$, ${\cal O}_{1,2}^{J/\psi \rho}$ and ${\cal O}_3^{DD^*}$, respectively.  Their energies are  almost equal to non-interacting energies $E^{n.i.}$ represented by the horizontal lines, which indicates that the interaction  in $I\!=\!1$ channel is small.   We find no extra state  in addition to the scattering states in the $I\!=\!1$ channel, thus no candidate state for $X(3872)$.

The sizable decay $X(3872)\to J/\psi \,\rho$ in experiment makes this state particularly interesting and gives rise to two popular interpretations. Both are based on the isospin breaking with the dominant effect coming from the   $8~$MeV isospin splitting of the $D^0\bar D^{0*}$ and $D^+\bar D^{-*}$ \cite{Tornqvist:2004qy,Gamermann:2009fv}. 
The first interpretation is based on $X(3872)$ with $I=0$ and the isospin is broken in the decay.  The second possibility is that $X(3872)$ is a linear combination  $|X(3872)\rangle=a_{I=0}|DD^*\rangle_{I=0}+a_{I=1}|DD^*\rangle_{I=1}$ with $a_{I=1}\ll a_{I=0}$ and  $a_{I=1}(m_u\!=\!m_d)\!=\!0$ \cite{Tornqvist:2004qy,Gamermann:2009fv}. Our non-observation  of $X(3872)$ with $I\!=\!1$  is in agreement with both interpretation due to exact isospin with $m_u\!=\!m_d$ in our simulation. Another possibility is that five $I\!=\!1$ scattering interpolators (\ref{O}) are not diverse enough to render $X(3872)$, which calls for simulations including also other types of $I=1$ interpolators in the future.   

\vspace{0.1cm}

{\it Results for I=0:} The lowest energy  state in Fig. \ref{fig:eff} is the conventional $\chi_{c1}(1P)$.  The energy state represented by the triangles  is $J/\psi(0)\omega(0)$; it disappears if   ${\cal O}^{J/\psi \omega}_{1,2}$ is not used in the basis of $C_{ij}(t)$, leaving the remaining energies and overlaps almost unmodified (Fig. \ref{fig:eff}b), which indicates that $J/\psi\, \omega$ is not significantly coupled to the rest of the system. 
The diamonds  correspond to $D(1)D^*(-1)$ and have largest overlap to ${\cal O}_3^{DD^*}$. 

There are two remaining levels (circles and stars) in  Figs. \ref{fig:eff}a and \ref{fig:eff}b and one of them has to be $D(0)\bar D^*(0)$. The other level  is an evidence for the presence of a physical state in the energy region near $D\bar D^*$ threshold  and we believe it is related to experimental $X(3872)$.  We emphasize again that no lattice simulation has found evidence for $X(3872)$ in addition to the scattering states yet.  

 One expects two possible interpretations of the energy levels with circles and stars  in  Figs. \ref{fig:eff}a,b,   but the quantitative analysis rules out the second option: 
\begin{enumerate}
\item
 Stars correspond to a weakly bound state $X(3872)$ slightly below $D\bar D^*$ threshold and circles correspond to the scattering state $D(0)\bar D^*(0)$, which is significantly shifted up due to a large negative $D\bar D^*$ scattering length $a_0^{DD^*}\equiv \lim_{p\to 0} \tan(p)/p$. 

 This is exactly a scenario envisaged for one shallow bound state on the lattice \cite{Sasaki:2006jn} and confirmed for deuteron in \cite{Yamazaki:2011nd,Beane:2013br}.  Levinson's theorem requires the $D\bar D^*$ phase shift to start at $\delta(p\!\!=\!\!0)\!\!=\!\!\pi$  and fall down to $\delta(p\!\to \!\infty)\!\!=0$ for  one near-threshold bound state. This implies negative $a_0^{DD^*}$ and positive energy shift of $D(0)\bar D^*(0)$  \cite{Luscher:1990ux}.  
 \item 
 The other possibility would be to identify the circles with a   resonance above $D\bar D^*$ threshold and stars with the down-shifted $D(0)\bar D^*(0)$ scattering level arising from the attractive interaction with positive $a_0^{D\bar D^*}$. This interpretation is however ruled out for our data which leads to $a^{DD*}_0<0$ in Eq. (\ref{a0_r0}) below. 
\end{enumerate}
We  conclude that $X(3872)$ is related to the energy level indicated by stars in Fig. \ref{fig:eff}a,b.

An interesting question is whether $X(3872)$ is accompanied slightly heavier state, sometimes called $\chi_{c1}(2P)$, with the same quantum numbers. Fig. \ref{fig:eff}a shows that we do not find a candidate for such a state for $E<4100~$MeV,  in agreement with experiment which also fails to find another $1^{++}$ state nearby. Our results therefore allow the possibility for interpreting $X(3872)$ as $\chi_{c1}(2P)=\bar cc$ accidentally aligned with $D\bar D^*$ threshold.  

We find $\chi_{c1}(1P)$  but no candidate for $X(3872)$ in $I=0$ channel if the interpolator basis consists only of five scattering interpolators (Fig. \ref{fig:eff}d). Perhaps this can be understood if $X(3872)$ is a consequence of accidental alignment of the  $c\bar c$ state with $D\bar D^*$ threshold, which may be absent in practice if ${\cal O}^{\bar cc}$ are not explicitly incorporated.  


\begin{table}[t]
\begin{ruledtabular}
\begin{tabular}{c|cccc}
level  & fit & $E_n\!-\!\tfrac{1}{4}(m_{\eta_c}\!+\!3m_{J/\psi})$ & $p^2$ & $p\cdot \cot \delta(p)$\\ 
  $n$ & $t$  & [MeV ]                        & [GeV$^2$] & [GeV] \\
\hline
1 & 6-11 & $429(3)$  & / & / \\
2 & 8-11 & $785(8)$  & $-0.075(15)$ & $-0.21(5)$ \\
3 & 6-9  & $946(11)$ & $0.231(22)$ &  $0.17(9)$\\
4 & 7-10 & $1028(18)$& / & /
\end{tabular}
\end{ruledtabular}
\caption{\label{tab:E}The energies extracted from the one-exponential correlated fit of the $6\times 6$  $C_{ij}(t)$ based on $O^{\bar cc}_{1,3,5},~{\cal O}^{DD^*}_{1,2,3}$ and $t_0\!=\!2$. The $p$ denotes $D$ and $D^*$ momentum and $\delta(p)$ denotes their scattering phase shift. }
\end{table}

The phase shifts $\delta(p)$ for the $s$-wave  $D\bar D^*$ scattering are extracted using the well-established and rigorous L\"uscher's relation \cite{Luscher:1990ux}
\begin{equation}
p\cdot \cot \delta(p)= \frac{2~Z_{00}(1;q^2)}{\sqrt{\pi}L}~,\qquad q^2\equiv \biggl(\frac{L}{2\pi}\biggr)^2 p^2 ~,
\end{equation}
which applies for elastic scattering below and above threshold. The $D$ and $\bar D^*$ momentum $p$ is extracted from  $E_{n}=E_D(p)+E_{D^*}(p)$ using dispersion relations $E_{D,D^*}(p)$ \cite{Mohler:2012na} and $E_{n=2,3}$  
from Fig. \ref{fig:eff}b and Table \ref{tab:E}. These energies result from  correlation matrix with ${\cal O}^{\bar cc},~{\cal O}^{DD^*}$ but without ${\cal O}^{J/\psi \omega}$, so we expect that  the effect of the $J/\psi \omega$ is negligible\footnote{Which is confirmed by comparing Figs. \ref{fig:eff}a and \ref{fig:eff}b.}.

The resulting $p\cot\delta$ in Table \ref{tab:E} for $p^2$ slightly below and above threshold can be described by the effective range approximation for $D\bar D^*$ scattering in $s$-wave
\begin{equation}
\label{eff_range}
p\cot \delta(p)=\frac{1}{a_0^{DD*}}+\frac{1}{2}r_0^{DD*} p^2~.
\end{equation}
Inserting $p\cot\delta(p)$ and $p^2$ for levels $n=2,3$ to (\ref{eff_range}), we get two relations which render the $(D\bar D^*)_{I=0}$ scattering length  and the effective range at our $m_\pi\simeq 266~$MeV 
\begin{equation}
\label{a0_r0}
a_0^{DD^*}=-1.7 \pm 0.4~\mathrm{fm}~,\quad r_0^{DD^*}=0.5\pm 0.1 ~\mathrm{fm}~.
\end{equation}

\begin{table}[t]
\begin{ruledtabular}
\begin{tabular}{c|cc}
$X(3872)$  &  $m_X\!-\!\tfrac{1}{4}(m_{\eta_c}\!+\!3m_{J/\psi})$ &  $m_X\!-\!(m_{D^0}\!+\!m_{D^{0*}})$ \\
\hline
lat$^{L\!\to\!\infty}$ &$815\pm 7~$MeV & $-11\pm 7~$MeV\\
exp                            & $804\pm 1$ MeV & $-0.14\pm 0.22~$MeV 
\end{tabular}
\end{ruledtabular}
\caption{\label{tab:X} $m_{X(3872)}$ from  lattice and experiment \cite{pdg12,Tomaradze:2012iz}.  }
\end{table}

The infinite volume $D\bar D^*$ bound state (BS) appears where S-matrix, $S\propto (\cot\delta(p)-i)^{-1}$, has a pole, so for the value of $p^2_{BS}<0$ where $\cot\delta(p_{BS})=i$.  The $D\bar D^*$ bound state $X(3872)$  appears near threshold, so we determine the binding momentum $p^2_{BS}=-0.020(13)~$GeV$^2$ which corresponds to $\cot\delta(p_{BS})=i$ from the effective range approximation (\ref{eff_range}) and the parameters (\ref{a0_r0}). This binding momentum then renders the position of the bound state $X(3872)$ in the infinite volume via 
$m_X^{lat}(L\!\!\to\!\! \infty)=E_D(p_{BS})+E_{D^*}(p_{BS})$  \footnote{We used  $D^{(*)}$ masses and dispersion relations  $E_{D^{(*)}}(p)$  from \cite{Mohler:2012na}, where they are extracted for employed configurations.  }
 and  the resulting mass in Table \ref{tab:X} is rather close to the experimental value. 

The errors  correspond to statistical errors based on  single-elimination jack-knife. The largest systematic uncertainty is expected from the finite volume corrections and we estimate that $m_X$ on the $DD^*$ threshold is also allowed 
  within our systematic errors, while $m_X$  above threshold is not supported due to $a_0^{DD*}\!<\!0$ (\ref{a0_r0}). Simulations on larger volumes will have to be performed to get more reliable  result for $m_X\!-\!m_D\!-\!m_{D^*}$, and the prospects are discussed in the Supplemental Material. 
The variation of $m_X$ with $m_\pi=[140,266]~$MeV is within this uncertainty according to the analytic study based on the molecular picture   \cite{Baru:2013rta}.  

Concerning the  composition of our candidate for $X(3872)$, Fig. \ref{fig:eff} shows its representative overlaps $\langle {\cal O}_i|n=2\rangle$. It has particularly sizable overlaps with $\bar cc$ and $D(0)D^*(0)$ interpolators, and has non-vanishing overlaps with the remaining ones. 
Note that the aim of the present paper was not to choose  between most popular interpretations ($\bar cc$ state accidentally aligned with $DD^*$ threshold or   $D\bar D^*$ molecule, etc. ), but rather to find a candidate for $X(3872)$ on the lattice and determine it's mass.

\vspace{0.1cm}

In conclusion, a candidate for  $X(3872)$ is found $11\pm 7~$MeV below the $D\bar D^*$ threshold using two-flavor dynamical lattice simulation  with $J^{PC}=1^{++}$ and $I\!=\!0$. In the simulation, the $X(3872)$ appears in addition to the nearby  $D\bar D^*$ and $J/\psi\,\omega$ discrete scattering states, and we extract large and negative $D\bar D^*$ scattering length.  We do not find a candidate for $X(3872)$ in the $I\!=\!1$ channel, which may be related to the exact isospin in  our simulation.    \\

{\bf Acknowledgments}

We thank Anna Hasenfratz for providing the gauge configurations and D. Mohler for providing the perambulators. We are grateful to D.$~$Mohler, E.$~$Oset, A.$~$Rusetsky, M.$~$Savage and in particular to C.B. Lang for insightful discussions on how to extract physics information from the energy levels in the presence of a bound state. We acknowledge the support by the Slovenian Research Agency ARRS project N1-0020 and by the Austrian Science Fund FWF project I1313-N27. S.P. acknowledges INT in Seattle. 
\bibliographystyle{h-physrev4}
\bibliography{Lgt}

\newpage
\section{ Supplemental material to ``Evidence for $X(3872)$ from $DD^*$ scattering on the lattice'' }

\normalsize

{\bf Wick contractions}

Figure 1 presents the Wick contractions that enter the correlation matrix $C_{ij}(t)$ (1) for the interpolators (2) in the main article. We evaluate all contractions from Fig. 1a and 1b. The contractions in Fig. 1b  involve charm annihilation and their effects on the charmonium(like) states are expected to be suppressed due to Okubo-Zweig-Iizuka rule. The results presented in the main article are based on all contractions in Fig. 1a ($a.1,~a.2,~...,a.13$). 

\vspace{0.2cm}

{\bf Finite volume effects and outlook}

We conclude by considering the finite volume effects in view of future prospects. The largest systematic uncertainty for a shallow bound state like $X$ is expected from the finite volume corrections and the leading effect was taken into account by calculating  $m_X^{L\!\to\! \infty}$ as shown in the main article. 
Simulations on larger volumes will have to be performed to get more reliable  result for $m_X\!-\!m_D\!-\!m_{D^*}$.  The relative uncertainty on $m_X\!-\!m_D\!-\!m_{D^*}$ is more challenging for the state that is closer to the threshold also because the spatial size of state is expected to be larger.  
Calculating $m_X\!-\!m_D\!-\!m_{D^*}$ with decent  relative uncertainty  will  definitely be challenging for a state which lies within $1~$MeV from the threshold experimentally.  Note that simulations on larger volumes will be more  challenging also since  the scattering states $D(p)D^*(-p)$ with $p\simeq \tfrac{2\pi}{L}|\mathbf{n}|$ will be spaced by less than $200~$MeV for $L>2~$fm and energy spectrum will be denser than in Fig. 1 of the main article.

\begin{figure*}[h!]
\begin{center}
\resizebox{0.87\textwidth}{!}{
\begin{tabular}{|c|c|c|c|}
\hline 
\includegraphics[width=0.2\textwidth]{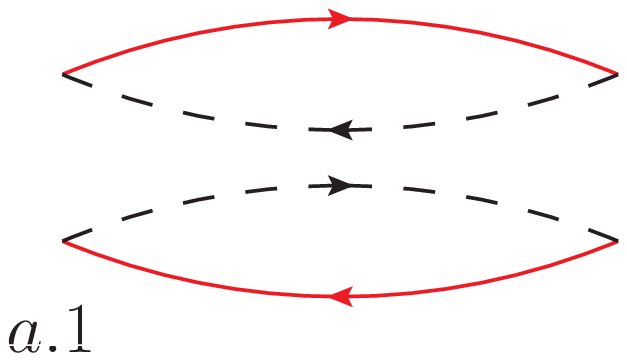}  & \includegraphics[width=0.2\textwidth]{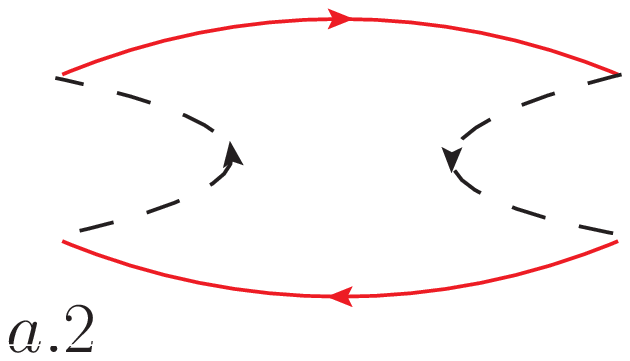}  & \includegraphics[width=0.2\textwidth]{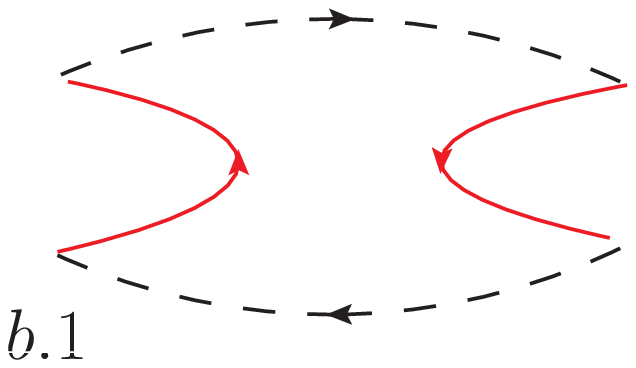}   &  \includegraphics[width=0.2\textwidth]{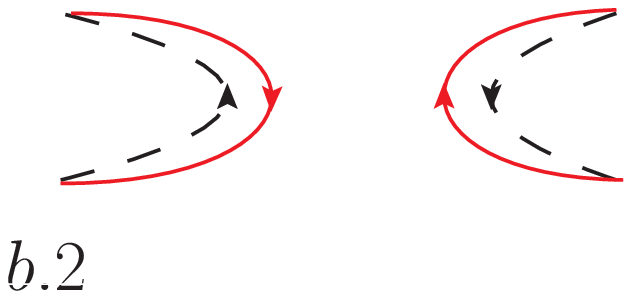}\\ 
\hline
\includegraphics[width=0.2\textwidth]{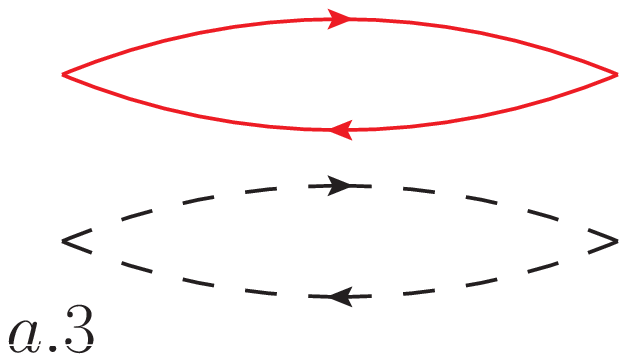}  & \includegraphics[width=0.2\textwidth]{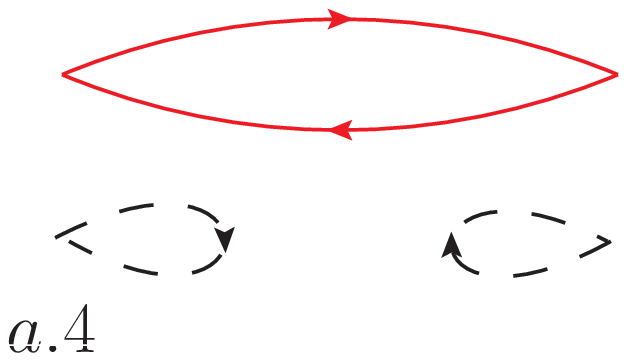} &  \includegraphics[width=0.194\textwidth]{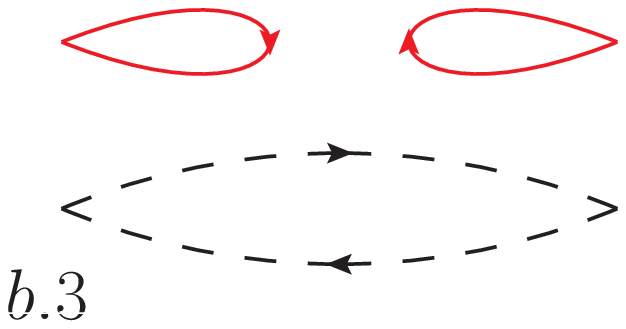} & \includegraphics[width=0.2\textwidth]{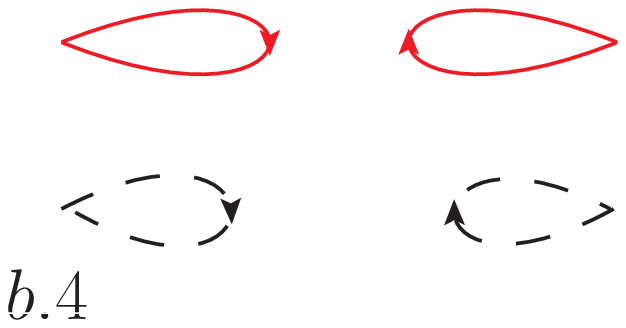} \\ 
\hline
\includegraphics[width=0.21\textwidth]{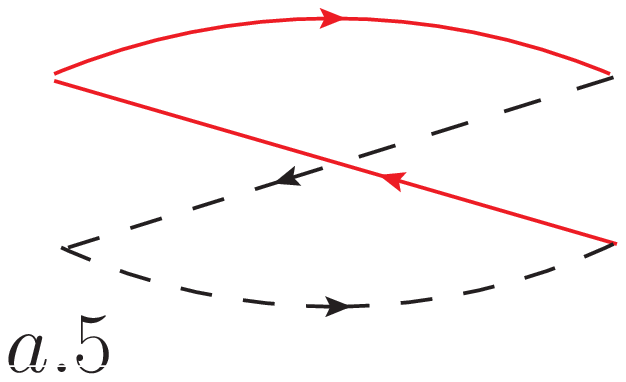}  & \includegraphics[width=0.2\textwidth]{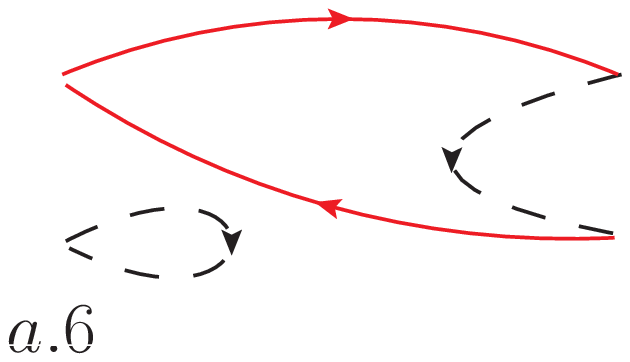}  &  \includegraphics[width=0.2\textwidth]{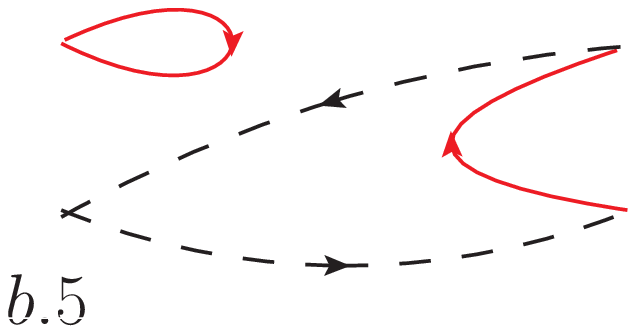}  & \includegraphics[width=0.2\textwidth]{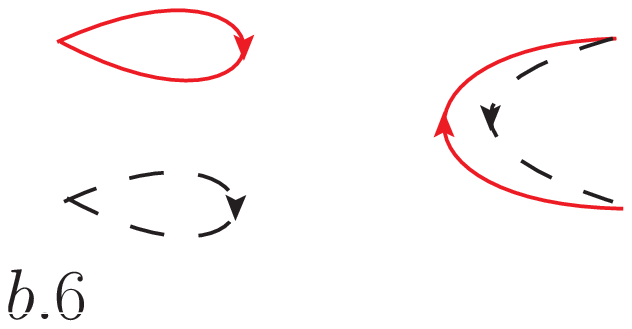} \\ 
\hline
\includegraphics[width=0.2\textwidth]{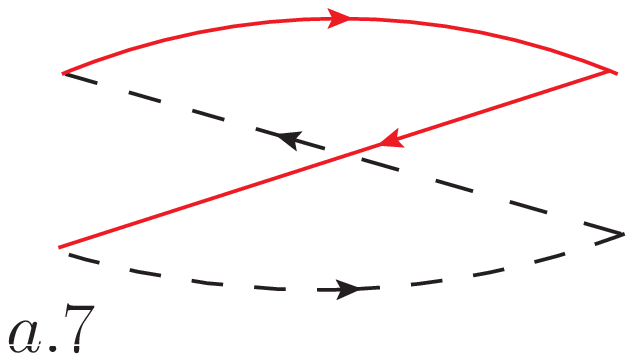}  & \includegraphics[width=0.2\textwidth]{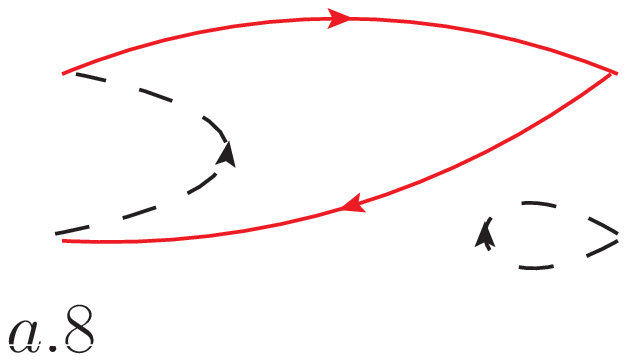}  &\includegraphics[width=0.2\textwidth]{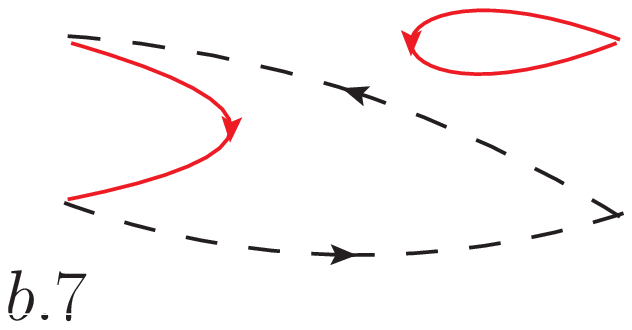}  & \includegraphics[width=0.2\textwidth]{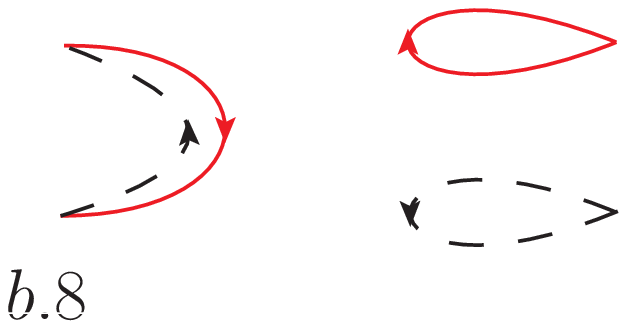} \\ 
\hline
\includegraphics[width=0.2\textwidth]{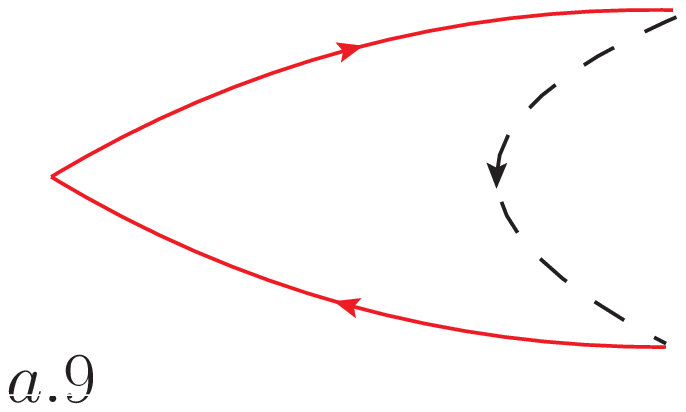}  & \includegraphics[width=0.2\textwidth]{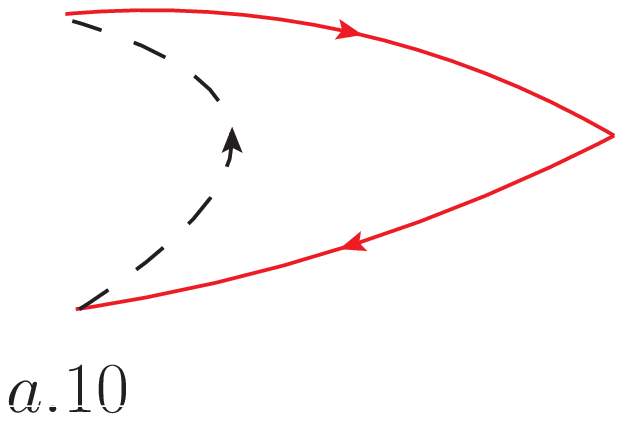}  & \includegraphics[width=0.2\textwidth]{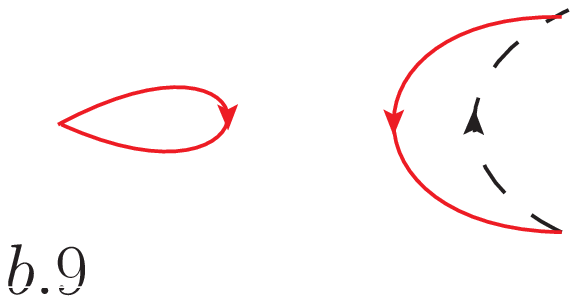}  &\includegraphics[width=0.2\textwidth]{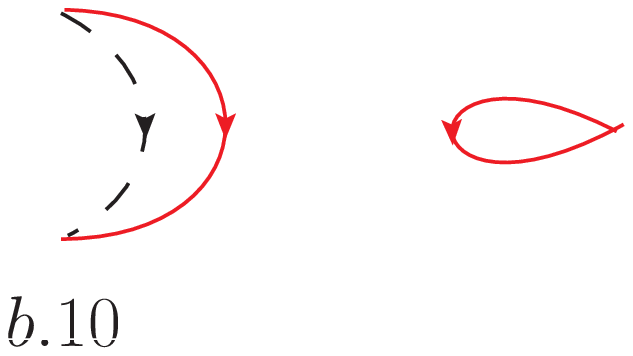} \\ 
\hline
\includegraphics[width=0.214\textwidth]{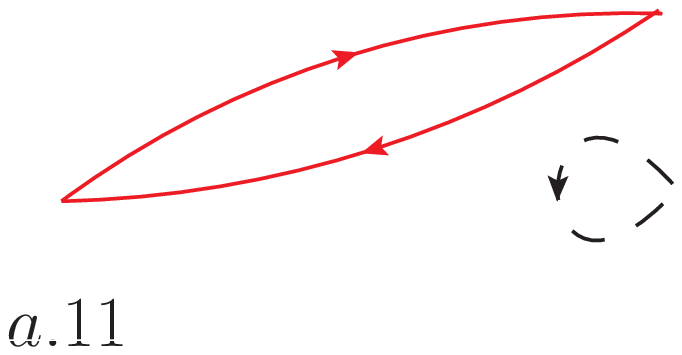}  & \includegraphics[width=0.2\textwidth]{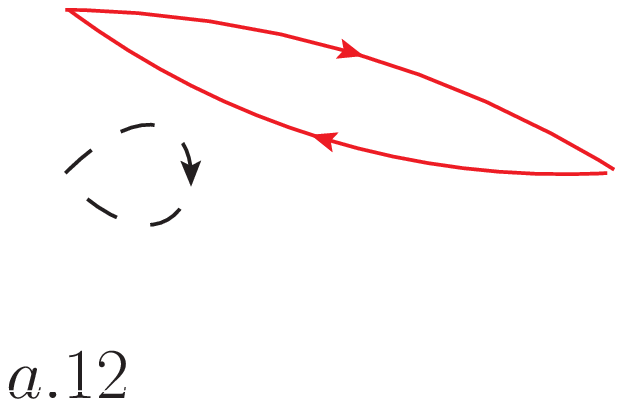} & \includegraphics[width=0.214\textwidth]{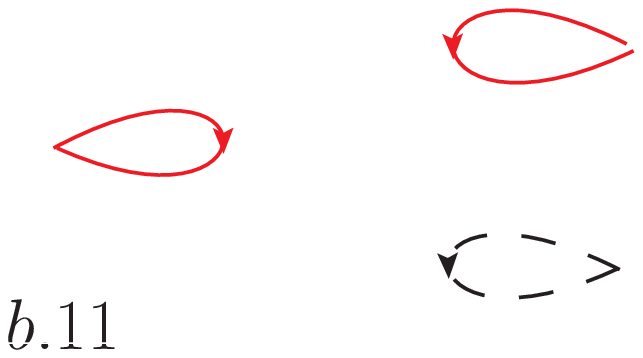} & \includegraphics[width=0.2\textwidth]{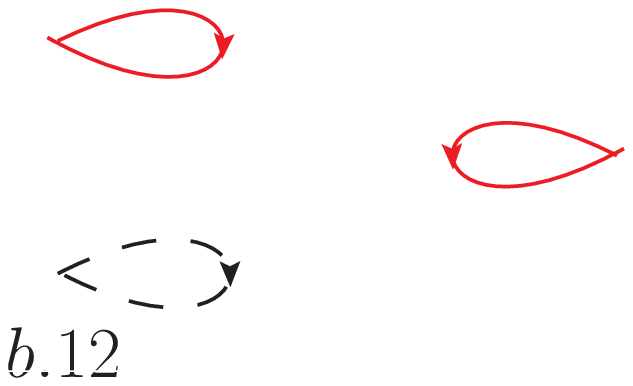}   \\ 
\hline
\includegraphics[width=0.2\textwidth]{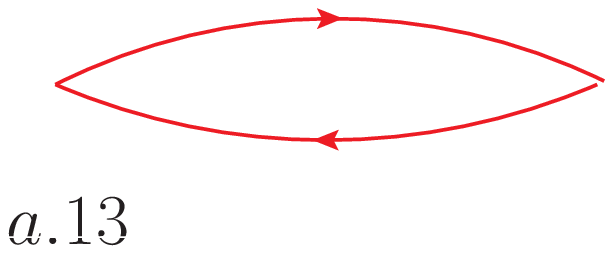}  & & \includegraphics[width=0.2\textwidth]{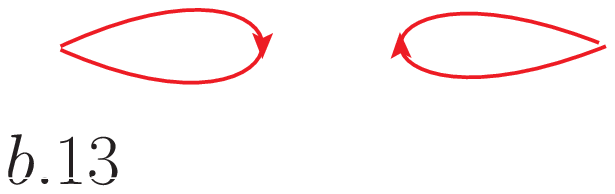} & \\ 
\hline 
\end{tabular} 
}
\end{center}
\caption{ Wick contractions that enter the correlation matrix $C_{ij}(t)$ (1) for the interpolators (2) in the main article. The red solid line represents $c$ quark, while black dashed line represents $u$ or $d$ quark.}
\end{figure*}

\end{document}